\documentclass{aa}  

\def\chandra{{\it Chandra\/}}

\def\heao1{{\it HEAO-1\/}}

\def\spitzer{{\it Spitzer\/}}

\def\xmm{{XMM-{\it Newton\/}}}

\def\isocam{{\it ISOCAM\/}}

\def\lsimeq{{_<\atop^{\sim}}}
\def\lesssim{\mathrel{\hbox{\rlap{\hbox{\lower4pt\hbox{$\sim$}}}\hbox{$<$}}}}
\def\gtrsim{\mathrel{\hbox{\rlap{\hbox{\lower4pt\hbox{$\sim$}}}\hbox{$>$}}}}
\newcommand{\cgs}{ ${\rm erg~cm}^{-2}~{\rm s}^{-1}$} 
\newcommand{\lum}{\rm erg~s$^{-1}$}

\def\xray{\hbox{X-ray}}

\usepackage{graphicx}
\usepackage{txfonts}
\usepackage{natbib}
\usepackage{rotating}

\begin{document}
%


\title{The HELLAS2XMM survey}

\subtitle{X. The bolometric output of luminous obscured quasars: \\ 
The Spitzer perspective}

\author{F. Pozzi,\inst{1,2}
        C. Vignali,\inst{1,2}
        A. Comastri,\inst{2}
        L. Pozzetti,\inst{2}
        M. Mignoli,\inst{2}
        C. Gruppioni,\inst{2}
        G. Zamorani,\inst{2}
        C. Lari,\inst{3}
        F. Civano,\inst{1,2}
        M. Brusa,\inst{4}
        F. Fiore,\inst{5}
        R. Maiolino,\inst{5}    
        \and 
        F. La Franca\inst{6}   
}


\institute{Dipartimento di Astronomia, Universit\`a degli Studi di Bologna, Via Ranzani 1, 
I--40127 Bologna, Italy
\and
INAF --- Osservatorio Astronomico di Bologna, Via Ranzani 1, I--40127 Bologna, Italy
\and
INAF --- Istituto di Radioastronomia (IRA), Via Gobetti 101, I--40129 Bologna, Italy
\and
Max Planck Institut f\"ur Extraterrestrische Physik (MPE), Giessenbachstrasse 1, 
D-85748 Garching bei M\"unchen, Germany  
\and
INAF --- Osservatorio Astronomico di Roma, Via Frascati 33, I--00040
Monteporzio-Catone (RM), Italy
\and
Dipartimento di Fisica,  Universit\`a degli Studi Roma Tre, Via della Vasca
Navale 84, I--00146 Roma, Italy
}

\authorrunning{F. Pozzi, C. Vignali, A. Comastri et al.}
\titlerunning{\spitzer\ observations of luminous obscured quasars}



\abstract
{}
{
We aim at estimating the spectral energy distributions (SEDs) and the physical
parameters related to the black holes harbored in eight high X-ray-to-optical (${F_X}/{F_{R}}>$10) 
obscured quasars at $z>0.9$ selected in the 2--10~keV band from the HELLAS2XMM survey.
}
{
We use IRAC and MIPS 24~$\mu$m observations, along with optical and $K_{s}$-band photometry, 
to obtain the SEDs of the sources. 
The observed SEDs are modeled using a combination of an elliptical template and torus emission 
(using the phenomenological templates of Silva et al. 2004) 
for six sources associated with passive galaxies; 
for two point-like sources, the empirical SEDs of red quasars are adopted. 
The bolometric luminosities and the $M_{BH}-L_{K}$ relation are used to provide an 
estimate of the masses and Eddington ratios of the black holes residing in these AGN.
}
{
All of our sources are detected in the IRAC and MIPS (at 24~$\mu$m) bands.
The SED modeling described above is in good agreement with the observed near- and 
mid-infrared data. 
The derived bolometric luminosities are in the range $\approx10^{45}-10^{47}$~\lum, and the 
median 2--10~keV bolometric correction is $\approx$25, consistent with the widely adopted 
value derived by Elvis et al. (1994).
For the objects with elliptical-like profiles in the $K_{s}$ band, we derive high stellar 
masses $(0.8-6.2){\times}10^{11}$~M$_{\odot}$, black hole masses in the range 
$(0.2-2.5){\times}10^{9}$~M$_{\odot}$, and Eddington ratios
$L/L_{Edd}<0.1$, suggesting a low-accretion phase.
}
{}


\maketitle

%

\section{Introduction}
\label{intro_sec}

Hard \xray\ surveys have clearly revealed the important role played by
obscured Active Galactic Nuclei (AGN) to reproduce the cosmic \xray\ background 
(XRB; e.g., \citealt{1995A&A...296....1C}) and have provided evidence that a significant fraction 
of the accretion-driven energy density in the Universe resides in obscured \xray\ sources 
(e.g., \citealt{2005AJ....129..578B}; \citealt{2006ApJS..163....1H}; 
\citealt{2006ApJ...645...95H}). 

\begin{table*}
\begin{center}
\begin{minipage}{12.3cm}
\caption{Properties of our targets}
\label{table1}
\begin{tabular}{l  c  c  c  c  c  c } \\\hline\hline 
Source Id. & 2--10~keV flux$^\dagger$       & $R$ & $K_{s}$   & Morph($K_{s}$)   & X/O & $z$  \\
           & ($10^{-14}\ $erg\ cm$^{-2}$\ s$^{-1}$)&     &    &   &   &      \\\hline
Abell~2690\#75      &    3.30   &   24.60   &   18.33   &  E &  1.86   & 2.13$^{a}$ \\
PKS~0312$-$77\#36   &    1.90   &   24.70   &   19.13   &  E &  1.66   &   -- \\
PKS~0537$-$28\#91   &    4.20   &   23.70   &   18.99   &  E &  1.60   &   -- \\
PKS~0537$-$28\#54   &    2.10   &   25.10   &   18.91   &  E &  1.86   &   -- \\
PKS~0537$-$28\#111  &    2.10   &   24.50   &   17.64   &  E &  1.62   &   -- \\
Abell~2690\#29      &    2.80   &   25.10   &   17.67   &  P &  1.99   & 2.08$^{b}$    \\
PKS~0312$-$77\#45   &    2.80   &   24.40   &   18.62   &  P &  1.71   &   -- \\
BPM~16274\#69       &    2.27   &   24.08   &   17.87   &  E &  1.49   & 1.35$^{b}$    \\
\hline
\end{tabular}
\end{minipage}
\vskip0.2cm\hskip0.7cm
\begin{minipage}[h]{12.6cm}
\footnotesize
$^\dagger$ 2-10 kev \xray\ fluxes from \cite{2004A&A...421..491P}. \\
$^{a}$ Tentative spectroscopic redshift from near-IR spectroscopic observations 
\citep{2006A&A...445..457M}. \\
$^{b}$ Spectroscopic redshift from near-IR spectroscopic observations \citep{2006A&A...445..457M}. \\
\end{minipage}
\end{center}
\end{table*}
%
Until recently, the limited information on the broad-band emission of the counterparts of 
obscured \xray\ sources prevented a reliable determination of their bolometric luminosities. 
The lack of a proper knowledge of the spectral energy distributions (SEDs) of
obscured sources has led many authors to adopt, in the computation of the bolometric luminosities, 
the average value derived by \citet{1994ApJS...95....1E}, although in that work the sample comprises 
mostly local unabsorbed quasars.
%
%
%
%
By the current work, we aim at providing a robust estimate of 
the bolometric luminosity for obscured sources, which is an essential parameter 
to derive the cosmic mass density of super-massive black holes (SMBHs, i.e., following the 
\citealt{1982MNRAS.200..115S} approach). 
A reliable estimate of the bolometric luminosity of obscured AGN is typically limited by the 
actual capabilities of disentangling the nuclear emission (related to the accretion processes) 
from that of the host galaxy which, unlike for unabsorbed quasars, often dominates at optical and 
near-infrared (near-IR) bands. 

A significant fraction of high-redshift, luminous obscured AGN (the so-called Type~2 quasars) 
may have escaped spectroscopic identification due to their faint optical counterparts, 
thus preventing current studies from an accurate sampling of obscured sources. 
%
%
Mid-infrared (mid-IR) observations appear to be fundamental for this class of objects, since 
they are only marginally affected by dust obscuration and are able to recover 
the nuclear emission. With mid-IR observations, we expect to reveal the nuclear 
radiation re-processed by the torus of the obscured active nuclei, 
which are often recognized as such by means of their \xray\ emission only. 
For these sources, the soft \xray\ emission, 
which is photo-electrically absorbed by the gas, and the optical emission, 
extincted by the dusty circumnuclear medium, are expected to be downgraded in 
energy and to emerge as thermally reprocessed radiation in the IR at wavelengths in the range 
between a few and a few hundred $\mu$m \citep{1997ApJ...486..147G}. 

The potentialities of mid-IR observations were firstly shown by 
\citet{2002A&A...383..838F}, who detected with \isocam\ at 15 $\mu$m about two-thirds 
of the \xray\ sources detected in the 5--10~keV band in the \xmm\ Lockman Hole survey. 
A similar high detection rate at 24~$\mu$m has been recently reported by 
\citet{2004ApJS..154..160R} and \citet{2005AJ....129.2074F} studying the \spitzer\ counterparts 
of the \chandra\ sources in the CDF-S and in the ELAIS-N1 field, respectively, 
within the SWIRE survey \citep{2004ApJS..154...54L}.

In this context, a new interesting class of objects is emerging from the current \xray\ 
surveys: these sources are characterized by a high ($>$1) X-ray-to-optical flux ratio 
(hereafter X/O)\footnote{X/O is defined as $\log{\frac{F_X}{F_R}}=\log{F_X}+\frac{R}{2.5}+5.5$.}; 
for comparison, unobscured Type~1 AGN have a broad distribution peaked at X/O$\approx$0. 
Objects with X/O $\gtrsim$1 are about 20\% of the hard \xray\ selected sources, 
and the fraction of these sources seems to remain constant over $\approx$~3 decades of \xray\ flux 
\citep{2004Ap&SS.294...63C}. By definition, sources with high X/O are among the faintest sources 
in the optical band. 
In the shallow, large-area \xray\ surveys (e.g., the HELLAS2XMM survey, 
with $F_{2-10\ keV}>10^{-14}$~\cgs\ over $\approx$~3~deg$^2$; \citealt{2002ApJ...564..190B}), 
where the identification of a sizable sample of sources with X/O$>$1 has been possible 
(e.g., \citealt{2003A&A...409...79F}), the X/O selection criterion has proven to be effective in 
selecting Type~2 quasars at high redshifts. 

We have performed a pilot program to study with \spitzer\ a sample of eight sources selected 
in the 2--10~keV band from the HELLAS2XMM survey on the basis of their high ($>$1) X/O and 
large column densities ($N_{\rm H}\ge$10$^{22}$~cm$^{-2}$). 
The sample observed with \spitzer\ has been previously investigated in other bands. 
The most surprising finding of the follow-up campaigns was the association of these sources with 
luminous near-IR objects \citep{2004A&A...418..827M}, placing them into the class of Extremely Red 
Objects (EROs, $R-K\ge$5). 

The outline of the paper is as follows: in Sect.~2 we present the sample 
selection; \spitzer\ data reduction and analysis are discussed in Sect.~3,
while in Sect. 4 we describe the analysis of the SEDs. 
Finally, in Sect.~5 we estimate the bolometric luminosities, the stellar masses of 
the host galaxies, the black hole masses, and the Eddington ratios. 

Throughout this paper we adopt the ``concordance'' (WMAP) cosmology 
($H_{0}$=70~km~s$^{-1}$~Mpc$^{-1}$, $\Omega_{\rm M}$=0.3, and 
$\Omega_{\Lambda}$=0.7; \citealt{2003ApJS..148..175S}). 
Magnitudes are expressed in the Vega system.

\section{Sample selection}
\label{sample_sel}

The eight objects presented in this paper (see Table~\ref{table1}) were selected among the 10 
HELLAS2XMM high X/O ratio sources detected in the $K_{s}$ band with ISAAC at ESO-VLT; 
for details on the association of the $K_{s}$-band counterpart to the \xray\ source, 
see \citet{2004A&A...418..827M}. 
Two sources of the original sample were not selected for \spitzer\ observations: 
one (PKS~0537$-$28\#31) is associated with a disky galaxy, 
while the other (BPM~16274\#181) has an ambiguous $K_{s}$-band morphological classification. 
All but one of the sources observed by \spitzer\ belong to the first square degree field 
(122 \xray\ sources; see \citealt{2003A&A...409...79F} for the spectroscopic and photometric 
identification and \citealt{2004A&A...421..491P} for the \xray\ spectral analysis); 
the only exception is BPM~16274\#69, which belongs to the sample of the second square degree 
(110 \xray\ sources; see \citealt{cocchia07}; Lanzuisi et al., in preparation). 

\begin{figure*}
\vglue-1.8cm
\centering
\includegraphics[width=\textwidth]{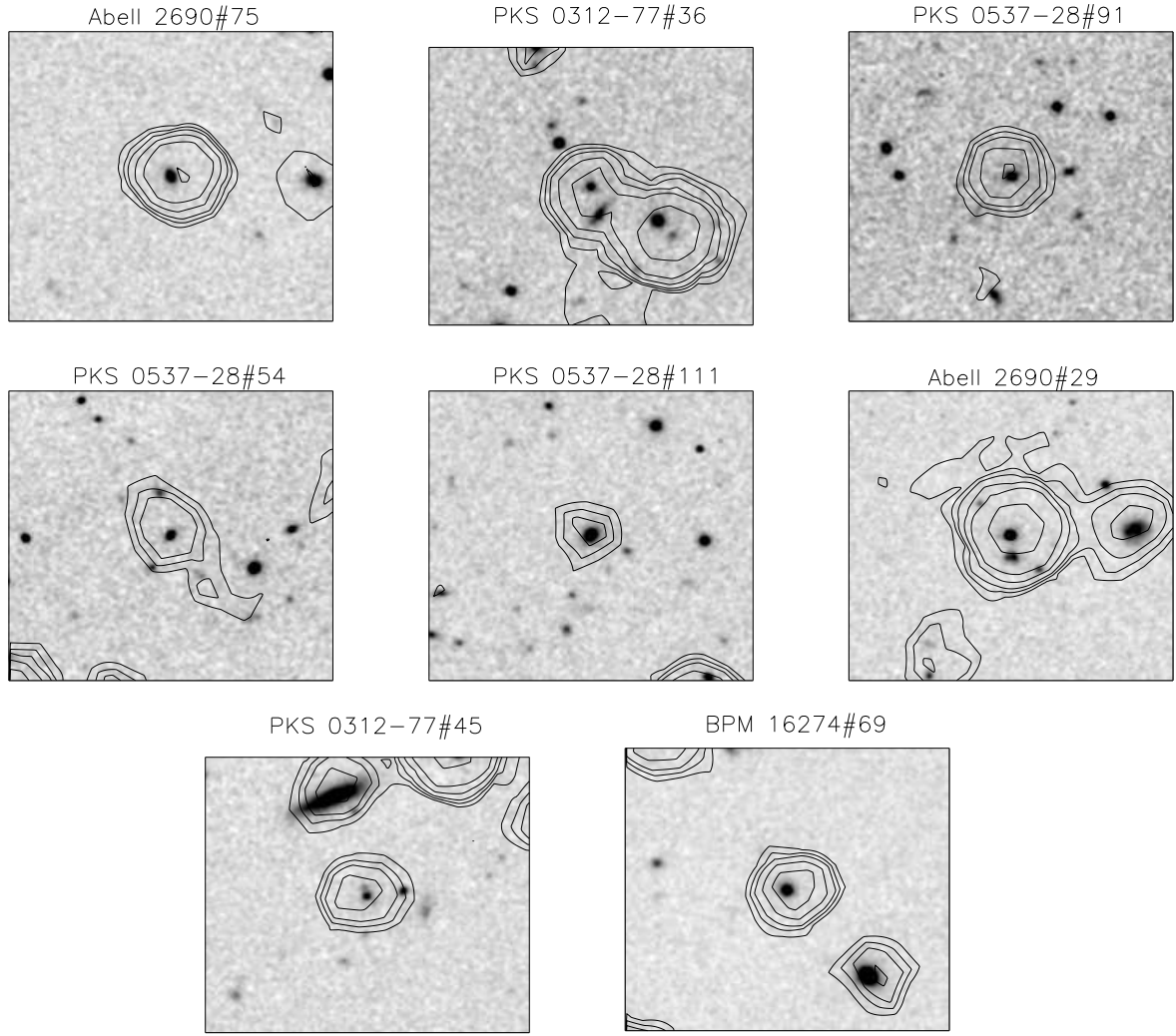}
\vglue-8cm
\caption{ISAAC $K_{s}$ images, centered on the $K_{s}$ counterpart of the \xray\ sources; 
         each box is 30$^{\prime\prime}$ wide. North is to the top and East to the left. 
         Contour levels of the 24~$\mu$m emission corresponding to [3, 4, 5, 7, 10, 20, 40]$\sigma$ 
         are superimposed to each image.}
\label{postage_fig}
\end{figure*}
The selected sources, although faint in the optical band (23.7$<$$R$$<$25.1), are all bright 
in the near-IR ($K_{s}{\lesssim}19$) and, all but one, have $R-K_{s}>5$, thus being EROs. 
\citet{2004A&A...418..827M} were able to study the surface brightness profiles of these sources 
in the $K_{s}$ band and obtained a morphological classification. 
Only two sources are classified as point-like objects, while all of the others are extended, 
with clear detection of the host galaxy and radial profiles consistent with those of elliptical 
galaxies. In this latter class of sources, the central AGN is evident in the
\xray\ band, while in the near-IR the host galaxy dominates. 
An upper limit to the contribution of a central unresolved source (i.e., the
nuclear emission), ranging from 2\% to 12\% of the galaxy emission, was obtained. 
Furthermore, using both the $R-K$ colour and the morphological information, 
a minimum photometric redshift, in the range 0.9--2.4, was estimated for these sources. 

For three of the sources with the reddest colours (Abell~2690\#75, Abell~2690\#29 and 
BPM~16274\#69; see Table~\ref{table1}), near-IR spectroscopic observations with ISAAC at 
ESO-VLT were performed by \citet{2006A&A...445..457M}, thus allowing for a spectroscopic 
identification for at least two of these sources (one redshift measurement appears tentative). 
The point-like source (Abell~2690$\#$29) shows the typical
rest-frame optical spectrum of high-redshift dust-reddened quasars,
with a broad H$\alpha$ line (\citealt{2002ApJ...564..133G}). 
The other two observed sources, both of them extended in the 
$K_s$ band, have narrow emission-line spectra: one is a LINER-like
object at $z$=1.35 and the second source has a spectrum with a
single weak line, tentatively associated with H$\alpha$ at $z$=2.13.
Consistently with the morphological information, in the first
source the AGN dominates the emission, while in the other two
sources the nuclear spectrum is heavily diluted by the host 
galaxy starlight.


\begin{table*}
\begin{minipage}{\textwidth}
\caption{\spitzer\ flux densities}
\label{table2}
\centering
\begin{tabular}{l  r   r   r   r   r   r } \\\hline\hline 
Source Id. & 3.6 $\mu$m    &   4.5 $\mu$m  & 5.8 $\mu$m &  8.0 $\mu$m &  24~$\mu$m       \\
           &        S$_{\nu}{\pm}{\Delta}$S$_{\nu}$     &    S$_{\nu}{\pm}{\Delta}$S$_{\nu}$   &   S$_{\nu}{\pm}{\Delta}$S$_{\nu}$
           &  S$_{\nu}{\pm}{\Delta}$S$_{\nu}$   &  S$_{\nu}{\pm}{\Delta}$S$_{\nu}$        \\\hline

Abell~2690\#75        &   51  $\pm$  5   &  56  $\pm$  6   &  89  $\pm$ 11  & 139 $\pm$ 15  &  565  $\pm$ 62 \\ 
PKS~0312$-$77\#36     &   41  $\pm$  4   &  44  $\pm$  5   &  40  $\pm$  8  &  71 $\pm$  9  &  236  $\pm$ 30$^a$ \\ 
PKS~0537$-$28\#91     &   28  $\pm$  4   &  35  $\pm$  4   &  42  $\pm$  8  &  80 $\pm$ 10  &  301  $\pm$ 40 \\ 
PKS~0537$-$28\#54     &   31  $\pm$  4   &  35  $\pm$  4   &  50  $\pm$ 10  &  47 $\pm$  8  &  279  $\pm$ 45 \\ 
PKS~0537$-$28\#111    &   88  $\pm$  9   &  75  $\pm$  8   &  41  $\pm$  6  &  46 $\pm$  7  &  148  $\pm$ 28 \\ 
Abell~2690\#29        &  141  $\pm$ 14   & 185  $\pm$ 19   & 260  $\pm$ 27  & 371 $\pm$ 38  & 1012  $\pm$106$^a$ \\ 
PKS~0312$-$77\#45     &   50  $\pm$  6   &  62  $\pm$  7   &  69  $\pm$ 10  &  78 $\pm$ 10  &  249  $\pm$ 35 \\ 
BPM~16274\#69         &   86  $\pm$  9   &  92  $\pm$  9   &  97  $\pm$ 11  & 120 $\pm$ 13  &  286  $\pm$ 34 \\ 

\hline
\end{tabular}
\end{minipage}
\vskip0.2cm\hskip3.6cm
\begin{minipage}[h]{11cm}
\footnotesize
The flux density is reported in units of $\mu$Jy. 
$^a$ The 24~$\mu$m flux density is probably over-estimated due to contamination from nearby 
sources and should be considered as an upper limit.
\end{minipage}
\end{table*}


\section{Spitzer observations and data reduction}
\label{reduction_sec}

The whole sample of eight hard \xray\ selected sources has been observed by
\spitzer\ \citep{2004ApJS..154....1W}, with IRAC \citep{2004ApJS..154...10F}
observations of 480~s integration time and MIPS \citep{2004ApJS..154...25R} 
observations at 24~$\mu$m with a total integration time of $\approx$~1400~s per position. 
IRAC observations were performed in photometry mode with frame time of 30~s
and dither pattern of 16 points. The MIPS 24~$\mu$m observations were 
performed in MIPS photometry mode with frame time of 10~s, 10 cycles and small-field pattern. 
To reduce overheads, the cluster option was used when possible. 

For the IRAC bands, we used the final combined post-basic calibrated data (BCD) mosaics produced 
by the \spitzer\ Science Center (SSC) pipeline (Version S12.0--S13.01). 
At 24~$\mu$m, we started the analysis from the BCD produced by the SSC pipeline 
(Version S12.4.2--S13.01) and then we applied {\it ad hoc} procedures to 
optimize the reduction, since some of our sources were close to the detection
limit (see Table \ref{table2}). 
We remind that BCD are individual frames already corrected for dark, flat field and geometric
distortion. We improved the quality of the BCD by correcting each individual BCD 
for a residual flat field depending on the scan mirror position \citep{2006AJ....131.2859F}. 
The residual flat fielding was obtained from our own data by averaging the BCD 
corresponding to the same scan mirror position and the same Astronomical Observation Request (AOR), 
considering all the different cluster positions. 
To each BCD, its median level was subtracted before this operation. 
This procedure was possible since our observations are not dominated by background fluctuations. 
The corrected BCD were co-added and background-subtracted using the SSC {\sc MOPEX} software 
\citep{2005PASP..117.1113M}. The resulting mosaics were made with
2.4$^{\prime\prime}$ pixel size. The overall analysis at 24 $\mu$m produces mosaics with a 
typical noise of $\approx$~0.020~MJy/pixel (a factor of 2 lower in comparison with the SSC pipeline 
mosaics). 
The noise map has been computed for each mosaic by scaling the measured mean
$rms$ of the central part of the map
according to the inverse square root of the coverage map. 

The flux densities of our targets in IRAC and MIPS bands were measured on the signal maps using 
aperture photometry at the position of the sources. 
The chosen aperture radius for the IRAC bands is 2.45$^{\prime\prime}$ and the
adopted factors for the aperture corrections are 1.21, 1.23, 1.38 and 1.58 (following the IRAC Data 
Handbook, Version 3.0) at 3.6 $\mu$m, 4.5 $\mu$m, 5.8 $\mu$m and 8 $\mu$m, respectively. \\
The chosen aperture radius at 24~$\mu$m is 7.5$^{\prime\prime}$; aperture
corrections were derived by examining the photometry of bright stars.  
Taking into account an additional correction of 1.15 to match the procedure used by 
the MIPS instrument team to derive calibration factors from standard star
observations, the resulting aperture correction is 1.57 (in agreement with 
the SWIRE team, see \citealt{surace05}). 

Table~\ref{table2} reports the results of the \spitzer\ observations. To
compute the photometric uncertainties, we added in quadrature the noise map
and the systematic uncertainties ($\approx$10\%, see MIPS and 
IRAC Data Handbook 2006, Version 3.0). The relative photometric 
uncertainties range from $\approx$10\% in the best cases (IRAC channels 1 and 2), 
up to $\approx$20\% at 24~$\mu$m in the worst ones.

All of the eight sources are clearly detected in both IRAC and MIPS 24~$\mu$m bands. 
At 24~$\mu$m, the flux densities span an order of magnitude, ranging between 
$\approx$1000~$\mu$Jy and $\approx$150~$\mu$Jy, with the faintest source (PKS~0537$-$28\#111) 
close to the 5$\sigma$ detection level. 



In Fig.~\ref{postage_fig} the $K_{s}$-band images, along with the contour levels 
of the 24~$\mu$m emission, are shown. At 24~$\mu$m, the sources PKS~0312$-$77\#36 and Abell~2690\#29 appear to be confused. In particular, in both cases, there is a second source
   at $\approx$~8-10\arcsec, unrelated to the targets. The contribution of these
   sources to the 24~$\mu$m flux density of our targets has been estimated by a
   decomposition analysis (using the PSF fitting algorithm IMFIT within
   the {\sc AIPS} environment).
  Furthermore, both the sources PKS~0312$-$77\#36 and Abell~2690\#29 present a
  second object at $\approx$~2\arcsec, clearly visible in the $K_{s}$ images, too close to our targets
  for a decomposition analysis, given the 24~$\mu$m pixel size. Since
  these close companions become increasingly fainter moving from the $K_{s}$
  bands to the longer IRAC wavelengths, we have attributed the entire flux density estimated
  from the decomposition analysis to our targets (see Table~\ref{table2}). 
  However, the 24~$\mu$m flux densities for these two sources should be
  treated as upper limits (see Fig.~\ref{silva_fig}).
%

%
We point out that the deblending procedure measures the peak flux 
density. To convert the peak flux density into total flux density, we have assumed that 
the deblended objects are point-like sources and applied a correction 
factor of 8.9 derived from the 24~$\mu$m \spitzer\ PSF and including the 1.15 calibration factor.

\section{Analysis of the spectral energy distributions}
\label{sed_sec}


In this section we provide an analysis of the SEDs of our sources, in order to derive the energy 
distribution of the nuclear component ``cleaned'' by the host galaxy contribution. 
As anticipated in Sect.~\ref{intro_sec}, the determination of the nuclear SEDs over a large 
wavelength range is an essential step to estimate the physical properties of the black hole, 
such as its bolometric luminosity, mass, and accretion rate. Taking advantage of the new \spitzer\ 
photometric points, simultaneously to the SED determination we have estimated the photometric 
redshifts of our sources. These new values are then compared with the minimum redshifts estimated by 
\citet{2004A&A...418..827M} using only the $R$ and $K_{s}$ bands and with the spectroscopic ones 
measured in three cases by \citet[see $\S$\ref{sample_sel}]{2006A&A...445..457M}.
  
Given the different morphological properties of our sources, two approaches have 
been adopted, one for the sources dominated in the $K_{s}$ band by the host galaxy 
(elliptical-like sources) and another one for the sources dominated by the nuclear 
component (point-like sources). 

\subsection{Elliptical-like sources}
\label{sed_elliptical_sec}

From the $K_{s}$-band morphological analysis, we know that at least up to the
observed 2.2 $\mu$m band the stellar contribution dominates the emission in these sources. 
At longer wavelengths, the nuclear component is expected to arise as reprocessed 
radiation of the primary emission, while the stellar component is expected to
drop (e.g., \citealt{2003MNRAS.344.1000B}; \citealt{2004MNRAS.355..973S}). 



In the analysis presented in this work, we have decided to follow a phenomenological approach, 
checking whether the emission of our sources can be reproduced as the sum of two components, one 
from the host galaxy and the other related to the reprocessing of the nuclear emission by the 
dusty torus envisaged by unification schemes \citep{1993ARA&A..31..473A}. 
The shape and relative strengths of the two components have to be consistent with all our 
observed data sets (multi-band photometry, $K_{s}$-band morphology and 
magnitude, $K_{s}$-band upper limit on the nuclear component, and \xray\ spectral analysis). 

\begin{figure*}
\centering
\includegraphics[width=\textwidth]{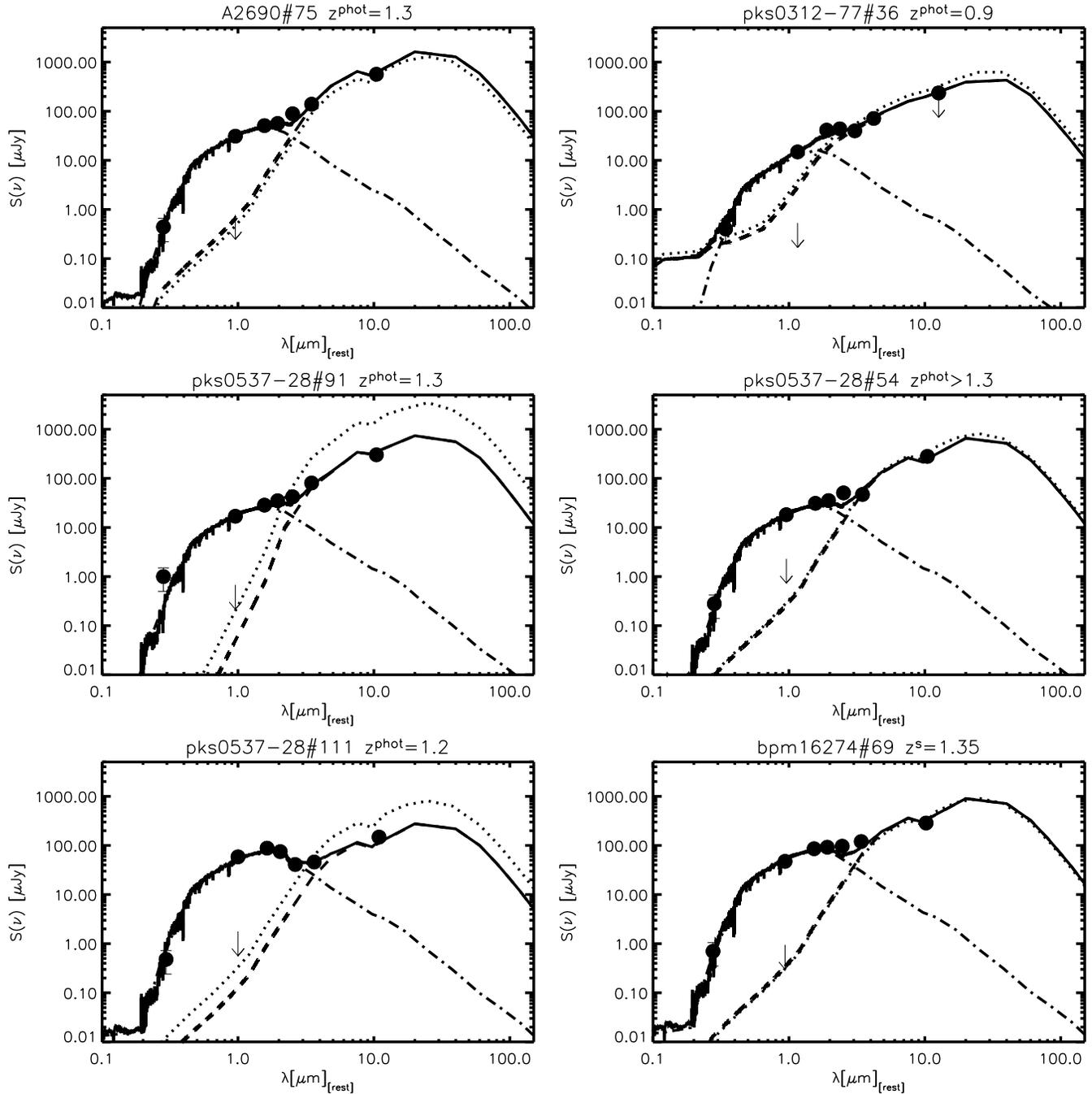}
\caption{Rest-frame SEDs of the elliptical sources (black filled circles) 
compared with the best-fit model obtained as the sum (solid line) of an early-type galaxy 
(dot-dashed line) and a nuclear component (dashed line). 
For comparison, the nuclear component as derived from the \xray\ normalization is 
also reported (dotted line). 
The nuclear $K_{s}$-band upper limits (downward-pointing arrows) were derived from 
the morphological analysis carried out by \citet{2004A&A...418..827M}. 
The 24~$\mu$m upper limit of source PKS~0312$-$77\#36 takes into account a possible 
contribution from a companion source. $z^{s}$ means that the redshift is spectroscopic 
(see $\S$\ref{sample_sel} for details), while $z^{phot}$ means that the redshift is photometric 
(see $\S$\ref{sed_elliptical_sec}).}
\label{silva_fig}
\end{figure*}

For the galaxy component, we adopted a set of six galaxy templates, obtained from the 
synthetic spectra of GISSEL~2003 \citep{2003MNRAS.344.1000B} assuming a simple 
stellar population and spanning a wide range of ages, from 1~Gyr up to a ``maximum age'' model 
($z_{form}=20$). 
The ``maximum age'' model has been adopted by \citet{2004A&A...418..827M} to derive the minimum 
photometric redshift for the sources of the current sample. 

For the nuclear component, we adopted the nuclear templates from 
\citet{2004MNRAS.355..973S}, based on the radiative transfer
models of \citet{1994MNRAS.268..235G}. We chose these templates since 
in the work of \citet{2004MNRAS.355..973S} the radiative transfer models are 
used to interpolate the observed nuclear IR data for a sizable sample of 
local AGN. We must note, however, that the nuclear 
observed SEDs are available only in the 2--20~$\mu$m regime, where data from small-aperture 
instruments are available. At wavelengths above $\approx$~20~$\mu$m, the SEDs are 
model extrapolations. 

\citet{2004MNRAS.355..973S} found that the nuclear SEDs can be 
expressed as a function of two parameters, the hard \xray\ (2--10~keV) intrinsic luminosity, 
which provides the normalization to the SED, and the column density N$_{H}$, which gives the shape 
to the SED (see Fig.~2 in \citealt{2004MNRAS.355..973S}). 
The shapes of the SEDs of the Seyfert galaxies are assumed to be valid also at quasar luminosities. 

In the attempt to provide a better estimate for the source redshifts, we used the four torus 
templates as given by \citet{2004MNRAS.355..973S}, which depend on the column densities N$_{H}$, and we left the normalizations free to vary. 
The redshift interval explored by this procedure is 0.5--3.0; 
the best-fit solution is obtained when the algorithm, based on the $\chi^{2}$, finds a minimum 
in the galaxy template, torus template and redshift parameter space. 

For four out of six sources the procedure finds a clear minimum $\chi^{2}$, which allows us to 
determine a photometric redshift with relatively good accuracy (see Table \ref{table3}). 
For sources BPM~16274\#69 and PKS~0537$-$28\#54, 
our procedure constrains only the lower bound of the redshift interval. For
all the six sources, 
the estimated redshifts are consistent with the minimum redshifts of \citet{2004A&A...418..827M}. 
In case of the source BPM~16274\#69, where a secure spectroscopic redshift is available, the 
minimum photometric redshift ($z_{phot}>1.25$) is consistent with the spectroscopic 
one ($z=1.35$). For source Abell~2690\#75, we find $z_{phot}=1.30^{+0.30}_{-0.20}$, which is
significantly lower than the spectroscopic value ($z=2.13$) reported by \citet{2006A&A...445..457M}. 
In this case, we choose to adopt the photometric determination, since the spectroscopic redshift 
is based on the tentative detection of a single line. 

The results of the SED fitting and decomposition are shown in Fig.~\ref{silva_fig}.
The dot-dashed line represents the best-fit galaxy template, the
dashed line is the best-fit nuclear template and the 
thick solid line is the sum of the two components. The best-fit galaxy 
templates are all typical of early-type galaxies with ages between 3 and 6 Gyr. 
We find an overall agreement between the SED templates and the data points. 
Moreover, in all but one of the sources, the nuclear component derived from the best fit 
is consistent with the upper limits derived from the analysis of the $K_{s}$-band images 
(shown as downward-pointing arrows). 
In Fig.~\ref{silva_fig} we also report as dotted line the SED of the nuclear component 
normalized to the intrinsic (i.e., de-absorbed) \xray\ luminosity following the prescriptions of 
\citet{2004MNRAS.355..973S}, where, at a given $N_{H}$, the normalization
depends only on the intrinsic hard X-ray luminosity.
The overall agreement between the SEDs normalized to the \xray\ 
luminosity and the best-fit SEDs is extremely interesting, being consistent within a factor of 
$\approx$~2--3. 

In Table \ref{table3} we report, along with the photometric redshifts, the column densities N$_H$ 
and the de-absorbed $L_{2-10\ keV}$ luminosities. 
We derive rest-frame N$_H$ column densities in the range $10^{22.0}$--$10^{23.4}$~cm$^{-2}$ and 
2--10~keV luminosities in the range \hbox{10$^{43.8}$--10$^{44.7}$}~\lum, placing these 
sources among the Type~2 quasar population.

\subsection{Point-like sources}
\label{pl_sources}

From the $K_{s}$-band morphological analysis, Abell~2690\#29 and PKS~0312$-$77\#45 
(see \citealt{2004A&A...418..827M} and Table~\ref{table1}) show a completely different appearance at 
2.2 $\mu$m in comparison with the sample of extended objects; 
these two sources have their near-IR emission mostly dominated by an unresolved source. 
The dominant role played by the AGN is supported for Abell~2690\#29 also by its near-IR spectrum, 
where a broad H${\alpha}$ emission line is detected (see \citealt{2006A&A...445..457M} and 
$\S$\ref{sample_sel}). Unfortunately, the near-IR spectroscopy information 
is absent for PKS~0312$-$77\#45 (see Table~\ref{table1}). 
\begin{figure*}
\centering
\includegraphics[width=\textwidth]{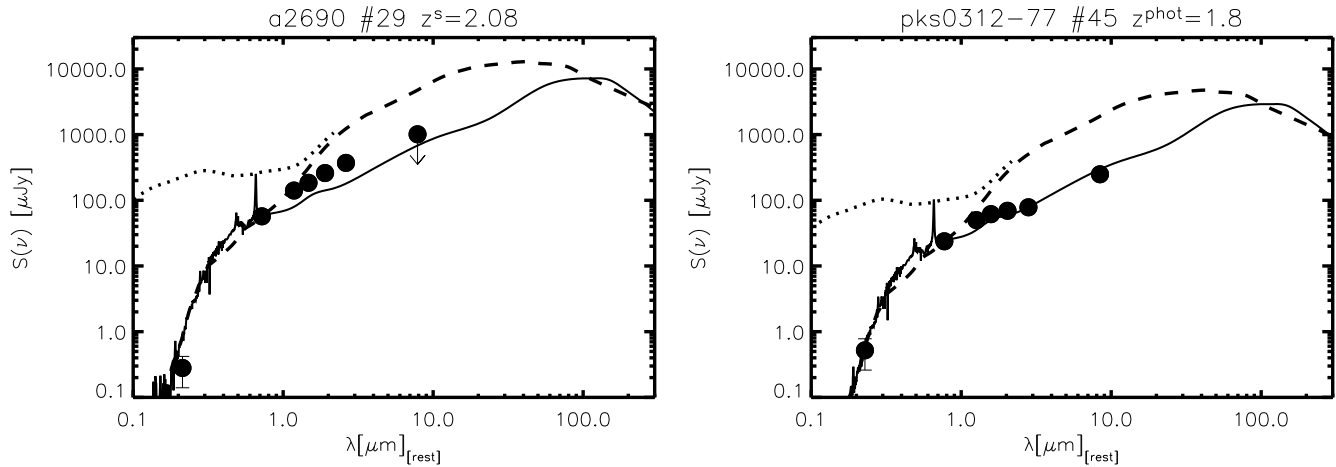}
\caption{Rest-frame SEDs of the two point-like sources (black filled circles) 
compared to an extinguished quasar template (dashed line) and the best-fit red quasar template 
(solid line). The extinguished quasar template is obtained from the unobscured quasar 
template of \citet{1994ApJS...95....1E} using the SMC extinction law with $E(B-V)=0.7$ and scaled 
to fit the $R-K_{s}$ colour (for comparison, the unobscured quasar template is also shown as 
a dotted line). 
The red quasar template is taken from \citet{2006ApJ...642..673P}. 
The 24~$\mu$m flux density upper limit for the source Abell~2690\#29 takes into account a possible 
contribution from a companion source. 
$z^{s}$ means that the redshift is spectroscopic (see $\S$\ref{sample_sel} for details), while 
$z^{phot}$ means that the redshift is photometric (see $\S$\ref{pl_sources}).} 
\label{pointlike_fig}
\end{figure*}

We first tried to reproduce their observed SEDs by reddening the composite template spectrum of 
bright Type~1 quasars of \citet{1994ApJS...95....1E} with several extinction laws. 
%
%
Reddening has been applied as prescribed by \citet{1997AIPC..408..403C} for a dust-screen model 
and by \citet{1992ApJ...395..130P} for the Small Magellanic Cloud (SMC) galaxy. 
The two prescriptions produce similar effects at ${\lambda}>0.5$ $\mu$m, but the SMC law produces 
redder spectra at shorter wavelengths for the same amount of extinction. 
Reddened templates with an SMC law reproduce quite well the optical spectra
of dust-reddened quasars in the Sloan Digital Sky Survey (SDSS; see
\citealt{2003AJ....126.1131R}), while, using the
\citet{1997AIPC..408..403C} law, \citet{2006ApJ...642..673P} were able to 
reproduce the SEDs of \xray\ sources in the \spitzer\ SWIRE survey. 


The procedure of reddening a typical Type~1 quasar does not provide a satisfactory fit to the 
photometric data points of our sources. In Fig.~\ref{pointlike_fig} we show the results 
obtained when the prescription of \citet{2004A&A...418..827M} for the extinction 
[SMC extinction law and $E(B-V)=0.7$] is adopted. 
   
The dashed line shows the reddened quasar template normalized to fit the $R-K_{s}$ colour. 
For comparison, the unobscured quasar template is also shown (dotted line). 
Although the $R-K_{s}$ colour is obviously reproduced, the overall SED is  
not well reproduced, since the observed IRAC and 24~$\mu$m flux densities are systematically 
lower than predicted (up to a factor of 10 at 24~$\mu$m). 

The discrepancy between the data points and the reddened Type~1 quasar
  template might be due to the application to active galactic nuclei of an
  extinction curve derived from galaxies. The different behaviour of
  AGNs from galaxies can be attributed to different dust distribution (i.e.,
  torus shape in AGNs), and gas-to-dust ratios, which can lead to an
  unusual dust reddening curve for AGNs.

Since a reddened quasar template does not reproduce the shape of our data, 
we adopted the red quasar template from \citet{2006ApJ...642..673P}, which is a composite spectrum: 
in the optical/near-IR band, 
it is the spectrum of the red quasar FIRST~J013435.7$-$093102 from \citet{2002ApJ...564..133G}, 
while the average of several bright quasars from the Palomar-Green (PG) sample 
\citep{1983ApJ...269..352S} with consistent optical data has been used in the IR. The
\citet{2006ApJ...642..673P} template reproduces the observed data points significantly better, 
allowing for the observed sharp decrease from 0.2 to 0.7 $\mu$m in the source rest frame 
(see Fig.~\ref{pointlike_fig} where the
\citealt{2006ApJ...642..673P} spectrum is shown as a solid line). 

For source Abell~2690\#29, which has a spectroscopic redshift, the SED normalization 
has been obtained through a best-fit procedure. For source PKS~0312$-$77\#45, 
where only a minimum redshift was available prior to this analysis, we have left free to vary both the
normalization and the redshift. 


In Table~\ref{table3} we report the derived redshifts, column densities N$_H$ and de-absorbed 
$L_{2-10\ keV}$ also for these sources which, similarly to the elliptical-like sources, 
belong to the Type~2 quasar population.

\section{Physical parameters}
\label{phys_param}
\subsection{Bolometric correction}

Once the SED of the nuclear component has been determined, the following step is to 
estimate the bolometric luminosity $L_{bol}$, which is a quantity directly related to
the central black hole activity. 

The bolometric luminosity $L_{bol}$ can be estimated from the luminosity in a given band
$b$, $L_{b}$, by applying a suitable bolometric correction $k_{bol,b}=L_{bol}/L_{b}$.
For the \xray\ selected sources, the bolometric luminosity is typically estimated 
from the luminosity in the 2--10~keV band
($k_{bol,2-10\ keV}=L_{bol}/L_{2-10\ keV}$).
In previous works, several authors used the 
bolometric correction obtained by \citet{1994ApJS...95....1E} for luminous, mostly nearby quasars, 
i.e., $k_{bol,2-10\ keV}{\approx}$~30. 
However, these corrections could be affected by the following uncertainties: firstly, they are 
average corrections obtained from a few dozens of bright quasars; 
secondly, as discussed in \citet{2004MNRAS.351..169M}, these corrections could overestimate 
the bolometric luminosities since they are based on the integral of the observed SEDs of 
bright unobscured AGN, without removing the IR bump (hence counting twice a fraction of 
$\approx$~30\% of the intrinsic optical--UV radiation). 
At lower luminosities (typical of Seyfert galaxies, i.e., $10^{42}-10^{44}$~\lum), a lower value 
for this correction ($k_{bol,2-10\ keV}{\approx}$~10) was suggested 
(e.g., \citealt{2004cbhg.symp..446F}). For heavily obscured luminous sources, only few objects 
have been studied in detail; in particular, for two SWIRE Compton-thick 
(i.e., log N$_H\gtrsim24$~cm$^{-2}$) AGN \citet{2006ApJ...642..673P} 
found $k_{bol,0.3-8\ keV}\approx$~3 and $\approx$~100. 

In this work, thanks to the multi-band observations and efforts in disentangling the AGN and the 
host components, we try to derive directly the nuclear bolometric luminosity of our sources 
without assuming any average correction. 
We estimate $L_{bol}$ by adding the \xray\ luminosity integrated over the 
entire \xray\ range ($L_{0.5-500\ keV}$, not corrected for absorption) to the IR luminosity 
($L_{1-1000\ {\mu}m}$). 

$L_{0.5-500\ keV}$ has been estimated from the observed $L_{2-10\ keV}$ luminosity 
assuming a single power-law spectrum with $\Gamma$=1.9 (typical for AGN emission; 
see, e.g., Fig.~6 of \citealt{2005AJ....129.2519V} and references therein) 
plus absorption (where the column densities are taken from the 
\xray\ spectral analysis; see \citealt{2004A&A...421..491P} and Lanzuisi et al., in preparation) 
and an exponential cut-off at 200~keV. The median value found for the ratio 
$L_{0.5-500\ keV}/L_{2-10\ keV}$ is ${\approx}$~4. 
The IR luminosity has been estimated by integrating the SED from 1~$\mu$m to
1000~$\mu$m using only the nuclear component for the AGN hosted in the elliptical galaxies
(see $\S$\ref{sed_elliptical_sec}), 
and the \citet{2006ApJ...642..673P} template for the point-like sources.

Before computing the bolometric output of our sources, 
the derived IR luminosities must be properly corrected to account for the 
geometry of the torus and its orientation. The first correction is related to the covering factor 
$f$ (which represents the fraction of the primary optical--UV radiation intercepted by the 
torus), while the second correction is due to the anisotropy of the IR emission, which is a function 
of the viewing angle (see \citealt{1993ApJ...418..673P} and \citealt{1994MNRAS.268..235G} 
for further details). 

We estimated the first correction ($\approx$~1.5) from the ratio of obscured 
(Compton thin + Compton thick) to unobscured quasars as required by the most recent \xray\ 
background synthesis model (see \citealt{2007A&A...463...79G}) in the luminosity range of our sources. 
A correction of $\approx$~1.5 implies an
average covering factor $f{\approx}0.67$ which, in a simple torus geometry, corresponds 
to an angle $\theta{\approx}48^{\circ}$ between the perpendicular to the equatorial plane 
and the edge of the torus. 

A first-order estimate of the anisotropy factor has been computed from the 
\citet{2004MNRAS.355..973S} templates as the ratio ($R$) of the luminosity of a face-on vs. an 
edge-on AGN, whose obscuration is parametrized as a function of N$_{\rm H}$. 
The integration has been performed in the 1--30~$\mu$m range, after normalizing the two SEDs to 
the same luminosity in the 30--100~$\mu$m range, where the anisotropy is thought to be negligible. 
The derived anisotropy factors are large only for the \citet{2004MNRAS.355..973S} template with 
higher column density ($R\approx$3--4 for N$_{\rm H}=10^{24.5}$~cm$^{-2}$); 
since all of our targets are characterized by lower obscuration, such corrections do not affect our 
IR luminosities significantly ($R\approx$1.2--1.3 for 
N$_{\rm  H}\approx10^{22.0}-10^{23.4}$~cm$^{-2}$). 
In conclusion, the final combined corrections to be applied to the observed IR 
luminosities of our sources, given their column densities, are in the range $\approx$~1.8--2.0. 
After adding the \xray\ luminosities, the IR correction factors would translate in a mean correction 
factor of $\approx$~1.7 in the computation of the bolometric luminosities.  


In Table \ref{table3} the derived bolometric luminosities are reported along with the full 
range ($L_{0.5-500\ keV}$) of \xray\ luminosities, the IR ($L_{1-1000\ {\mu}m}$) 
luminosities and the bolometric corrections ($k_{bol,2-10\ keV}$). We
  note that our $L_{1-1000\ {\mu}m}$ estimates (hence $L_{bol}$) are robust
  despite the choice of our SEDs. By comparing the $L_{1-1000\
    {\mu}m}$ obtained using the \citet{2004MNRAS.355..973S} model with the $L_{1-1000\ {\mu}m}$
    obtained adopting other recent average quasar SEDs
    (i.e., \citealt{2006ApJS..166..470R}), we have verified that the uncertainties in
    $L_{1-1000\ {\mu}m}$ are within the $\approx$~10\% level.
%
%
The bolometric output of our targets is dominated by the IR reprocessed emission, 
the primary \xray\ radiation ($L_{0.5-500\ keV}$) accounting only for $\lesssim$15\% of 
the total luminosity. 

\begin{figure}
\centering
\includegraphics[width=9cm]{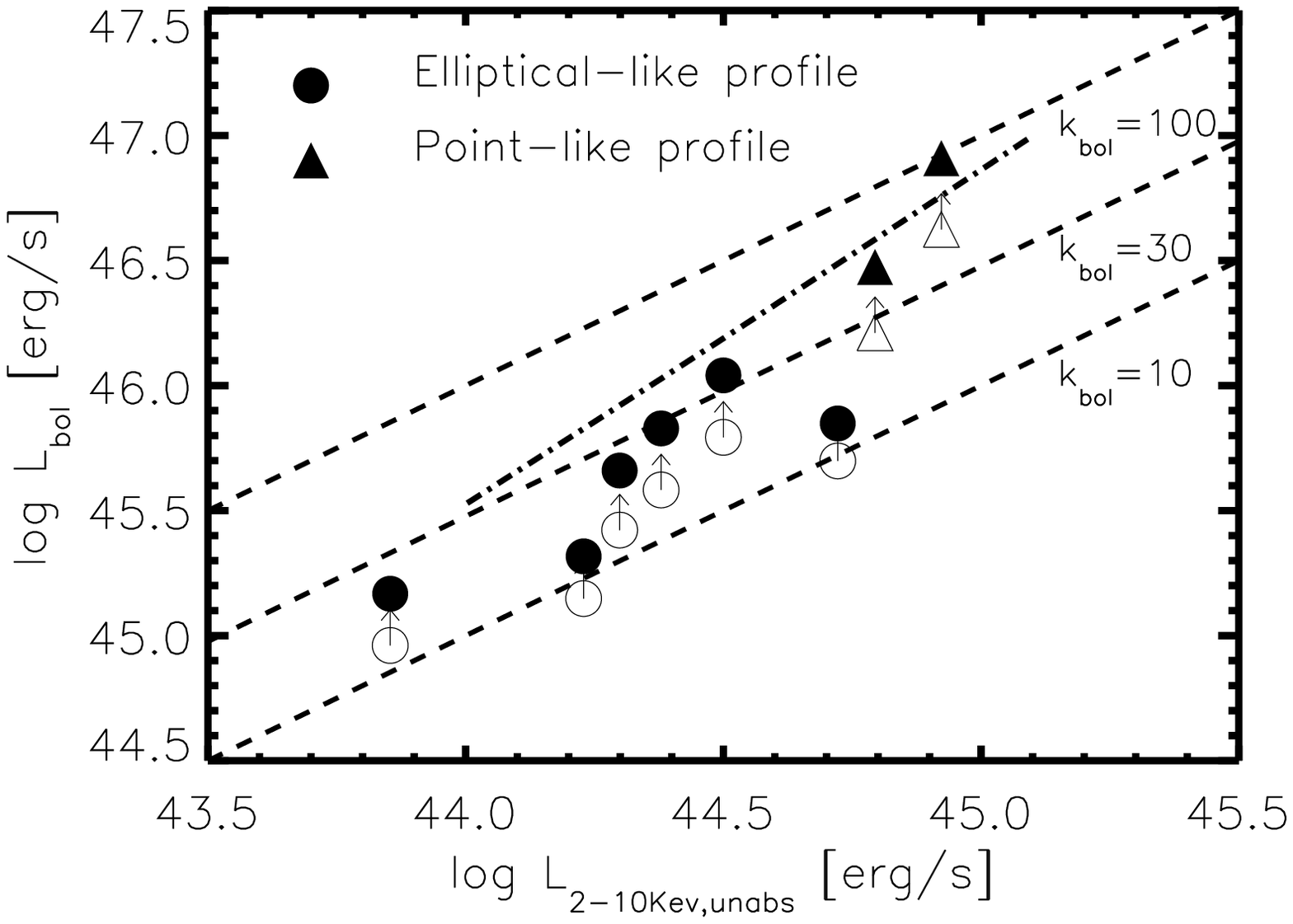} 
\caption{Bolometric luminosity vs. absorption-corrected 2--10~keV luminosity 
for the six AGN hosted in the elliptical galaxies (circles) and the two point-like AGN (triangles). 
The filled symbols refer to the values corrected for the covering factor and torus anisotropy, while 
the empty symbols refer to the bolometric luminosity without applying these corrections. 
The dot-dashed line represents the correlation from \citet{2004MNRAS.351..169M}. 
The three dashed lines represent the loci of $k_{bol,x}$=10, 30 and 100.}
\label{lx_lbol_fig}
\end{figure}

The derived median (mean) value of $k_{bol,2-10\ keV}$ is $\approx$~25 ($35\pm{9}$; 
see Fig.~\ref{lx_lbol_fig} and column 8 of Table~\ref{table3}), 
consistent with the value $k_{bol,2-10\  keV}{\approx}$~30 from 
\citet{1994ApJS...95....1E} widely adopted in past works. 
However, as pointed out also by \citet{1994ApJS...95....1E}, the bolometric corrections span 
a wide range of values ($\approx$~12--100); as a consequence, the adoption of a mean 
value could lead to inaccurate results.

We note that in the extreme case where no corrections for the covering factor and the anisotropy 
of the torus are applied, we would obtain a median $k_{bol,2-10\ keV}$ of ${\approx}$~16. 


In Fig.~\ref{lx_lbol_fig} the derived $L_{bol}$ as a function of $L_{2-10\ keV}$ is shown; 
the dot-dashed line joins the expected values from the analysis of \citet{2004MNRAS.351..169M}, 
where $k_{bol,2-10\ keV}$ is derived by constructing an AGN reference template taking into account 
how the spectral index $\alpha_{ox}$ \citep{1981ApJ...245..357Z} varies as a function of the 
luminosity \citep{2003AJ....125..433V}. 
Although the bolometric luminosities estimated for our objects are on average 
lower than those expected on the basis of \citet{2004MNRAS.351..169M} 
relation, they are however consistent with a trend of higher $k_{bol,2-10\  keV}$ 
for objects with higher \xray\ luminosity. If we fit our objects in the $L_{bol}-L_{2-10keV}$ plane 
with the same slope as the \citet{2004MNRAS.351..169M} relation, 
the difference in normalization is $\approx$~50\%. 

\begin{figure}
\centering
\includegraphics[width=9cm]{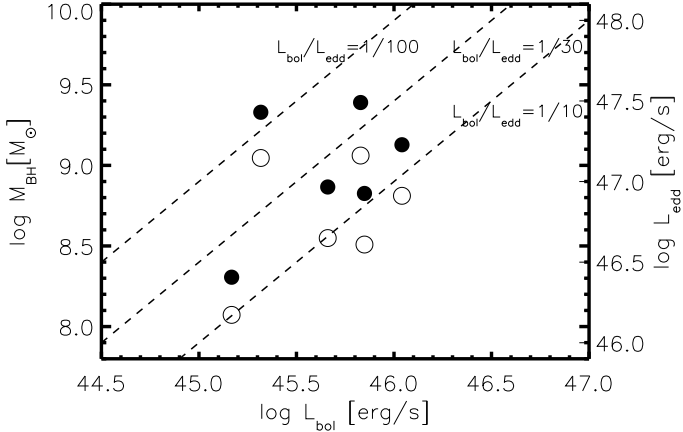}
\caption{Black hole mass ($M_{BH}$) vs. bolometric luminosity ($L_{bol}$) for the 
AGN hosted in the elliptical galaxies. The Eddington luminosity for a given
$M_{BH}$ is reported in the right-hand axis.  
The black hole masses have been estimated from the local $L_K-M_{BH}$ 
relation (\citealt{2003ApJ...589L..21M}) under two different hypotheses (see \ref{masses_sec}): 
(1) evolution by a factor of two of the $M_{BH}/M_{star}$ ratio with redshift in comparison to the 
local values (black filled circles); (2) no evolution of the $M_{BH}/M_{star}$
ratio (empty circles). 
The three dashed lines represent the loci of $L_{bol}/L_{Edd}=$ 0.01, 0.033 and 0.1 (from left to 
right).}
\label{ledd_fig}
\end{figure}

\begin{sidewaystable*}
\caption{Inferred rest-frame properties of our targets}
\label{table3}
\centering
\begin{tabular}{l  c  c  c  c  c  c  c  c  c  c  c} \\\hline\hline
Source Id.  &  z$^a$  & $N_H^b$   &  $L_{2-10\ keV}^c$  &
$L_{0.5-500\ keV}^d$   &$L_{1-1000\ {\mu}m}^e$ & $L_{bol}$ &
$L_{bol}/L_{2-10\ keV}$&  $L_{K}^f$  & $M_{star}$   &    $M_{BH}^g$  & 
($L_{bol}/L_{Edd}$)$^h$ \\
       &      & (10$^{22}$~cm$^{-2}$)  &  (10$^{44}$~erg~s$^{-1}$)  & ($10^{45}$~erg~s$^{-1}$) & 
($10^{45}$~erg~s$^{-1}$) &     ($10^{45}$~erg~s$^{-1}$) & &
($10^{11}L_{k,\odot}$) & ($10^{11}$M$_{\odot}$) & ($10^{9}$M$_{\odot}$)  \\\hline

Abell~2690\#75   &  1.30$^{+0.30}_{-0.20}$ &  6.9 &  3.2 &  1.3 &  9.7 & 11.0 & 34.7  & 5.2 &  3.5 &  1.3  &0.065\\
PKS~0312-77\#36  &  0.90$^{+0.05}_{-0.15}$ &  1.0 &  0.7 &  0.2 &  1.2 &  1.5 & 20.6  & 1.0 &  0.8 &  0.2  &0.058\\
PKS~0537-28\#91  &  1.30$^{+0.40}_{-0.70}$ & 25.8 &  5.3 &  2.6 &  4.4 &  7.1 & 13.4  & 2.8 &  1.5 &  0.7  &0.084\\
PKS~0537-28\#54  &  $>$1.30                &  1.6 &  2.0 &  0.6 &  3.9 &  4.6 & 23.0  & 3.0 &  2.0 &  0.7  &0.049\\
PKS~0537-28\#111 &  1.20$^{+0.20}_{-0.10}$ &  9.1 &  1.7 &  0.7 &  1.4 &  2.1 & 12.3  & 7.8 &  6.2 &  2.1  &0.008\\
Abell~2690\#29   &  2.08                   &  2.1 &  8.4 &  2.8 & 78.4 & 81.2 & 97.0  & 4.17 &  -- &  --   & --\\
PKS~0312-77\#45  &  1.85$^{+0.20}_{-0.30}$ &  8.0 &  6.2 &  2.6 & 27.2 & 29.8 & 47.9  & 1.65 &  -- &  --   &--\\
BPM~16274\#69    &  1.35                   &  2.5 &  2.4 &  0.8 &  5.9 &  6.7 & 28.2  & 8.8 &  6.0 &  2.5  &0.022\\\hline

\hline
\end{tabular}
\vskip0.2cm
\begin{minipage}[h]{0.96\textheight}
\footnotesize
$^a$ Photometric redshifts as derived from the analysis presented in this paper 
(see $\S$\ref{sed_elliptical_sec}). 
For source PKS~0537-28\#54, only a minimun redshift was estimated; 
for sources Abell~2690\#29 and BPM~16274\#69, the spectroscopic redshifts measured by 
\citet{2006A&A...445..457M} are reported. \\ 
$^b$ The column densities, measured through \xray\
  spectral fitting (see \citealt{2004A&A...421..491P} and Lanzuisi et al., in
  preparation, for details), were ``matched'' to the
 redshift used in the SED best-fitting procedure (using the relation 
 N$_{H}(z)$=N$_{H}(z=0){\times}(1+z)^{2.6}$). \\
$^c$ Absorption-corrected \xray\ luminosity. \\
$^d$ The 0.5--500~keV luminosities have been derived from the observed 2--10~keV 
luminosities as described in the text. The luminosities are not corrected for absorption.\\
$^e$ The 1--1000~$\mu$m luminosities have been derived from the integral of the nuclear SEDs 
(including the corrections described in the text). The values reported for the point-like sources refer to the 
\citet{2006ApJ...642..673P} red quasar template (see \ref{pl_sources} for details). \\
$^f$ The $K_{s}$-band luminosities refer to the nuclear component for the point-like sources 
(Abell~2690\#29 and PKS~0312$-$77\#45) and to the host-galaxy starlight for the AGN hosted in the 
elliptical galaxies. \\
$^{g,h}$ The reported $M_{BH}$ and $L_{bol}/L_{Edd}$ have been computed from the 
local $L_K-M_{BH}$ relation (\citealt{2003ApJ...589L..21M}), under the hypothesis of an evolution 
of the $M_{BH}/M_{star}$ ratio of a factor two with redshift (see $\S$\ref{masses_sec} for details). 
\end{minipage}
\end{sidewaystable*}

\subsection{Galaxy and black hole masses, and black hole Eddington ratios}
\label{masses_sec}

For the AGN hosted in elliptical galaxies we are able to infer both the galaxy and the black hole
masses. The galaxy masses are estimated, assuming a \citet{1955ApJ...121..161S} initial mass
function (IMF), 
from the $K_{s}$ luminosities taking 
into account that $M_{star}/L_{K}$ for an old stellar population can vary from
$\approx$~0.5 to $\approx$~0.9 (for ages between 3 and 6 Gyr;
\citealt{2003MNRAS.344.1000B}). 
We can derive $M_{star}$ directly from $L_{K}$ since for these sources the $K_{s}$-band emission 
is dominated by the galaxy starlight. The rest-frame $L_{K}$ have been derived using the 
appropriate SED templates (see Sect.~\ref{sed_elliptical_sec}). 
The inferred stellar masses are in the range \hbox{(0.8--6.2)$\times10^{11}$~$M_{\odot}$}, 
implying that our obscured AGN are hosted by massive elliptical galaxies at high redshifts. 
In Table~\ref{table3} both the $L_{K}$ and $M_{star}$ values are reported; we
note that the different assumption of the \citet{2003PASP..115..763C} IMF would produce a factor
$\approx$~1.7 lower masses \citep{2005A&A...442..125D}.

To estimate the black hole masses, we take advantage of the local
$M_{BH}-L_{K}$ relation \citep{2003ApJ...589L..21M} which, taking into account
the $M_{star}/L_{K}$ values, is expression of the intrinsic $M_{BH}-M_{star}$ relation. 
Given the challenging measurements of high-redshift black hole masses, the behaviour of 
this relation with redshift is still matter of debate and different authors, using different 
techniques, have found different results.  

\citet{2006ApJ...645..900W} and \citet{2006ApJ...649..616P} derive a significant 
evolution of the $M_{BH}-M_{star}$ relation with redshift, being the $M_{BH}/M_{star}$ ratio larger, 
at high redshift, up to a factor $\approx$~4 in comparison to the local value. 
In the \citet{2006ApJ...645..900W} analysis the discrepancy with respect to
the local value is already present at $z=0.36$, while
\citet{2006ApJ...649..616P} find an average $M_{BH}/M_{star}$ a factor
$\gtrsim$4 times larger than the local value at $z>1.7$, while at lower redshifts 
($1\lesssim z\ \lesssim 1.7$) they derive a ratio which is at most two times
higher than the local value, and maybe consistent with marginal or no evolution. 
On the other hand, \citet{2006NewAR..50..809S} and \citet{2006ApJS..163....1H} suggest that the 
$M_{BH}/M_{star}$ ratio is not significantly higher (at most a factor of two) than that 
measured locally up to $z\lesssim2$. 


Given these uncertainties about the evolution of the $M_{BH}-M_{star}$ relation with redshift, 
we have estimated the black hole masses for our objects under two different hypotheses: 
(1) the $M_{BH}/M_{star}$ ratio is higher than locally by a factor two in the 
redshift range ($0.9\lesssim z \lesssim 1.4$) of our sources; 
(2) the $M_{BH}/M_{star}$ ratio does not evolve with redshift.

In both cases, our results imply very massive black hole masses 
(see Fig.~\ref{ledd_fig}), in the range $\approx2.0\times10^{8}-2.5\times10^{9}$~$M_{\odot}$
in the former hyphothesis, with a factor of two lower values under the second hypothesis.

Our estimated $M_{BH}$ are consistent with the results derived by \citet{2004MNRAS.352.1390M} 
studying a large sample of Type~1 SDSS quasars and deriving the black hole masses 
from virial methods; most of their black hole masses are in the range 
1.5$\times10^{8}$--2.5$\times10^{9}$~$M_{\odot}$ in the redshift interval of our sample 
(see Fig.~1 of \citealt{2004MNRAS.352.1390M}). 

From the comparison of the bolometric luminosities computed in the previous section 
(see Table~\ref{table3}) with the Eddington luminosities calculated from the black hole masses 
estimated above, we derive that our obscured AGN are radiating at a relatively
low fraction of their 
Eddington luminosity (\hbox{$\lambda\approx$~0.008--0.084} and 
\hbox{$\lambda\approx$~0.015--0.170} under the two hypotheses; see Fig.~\ref{ledd_fig} and 
Table~\ref{table3}). This finding confirms and extends to a larger sample the results found by 
\citet{2006A&A...445..457M} for two sources of our sample and by \cite{2005A&A...432...69B} 
for a sample of EROs in the ``Daddi Field''. 
As suggested by \citet{2006A&A...445..457M}, the data indicate that our very massive black 
holes may have already passed their rapidly accreting phase and are reaching their 
final masses at low accretion rates. 
%

\begin{figure}
\centering
\includegraphics[width=9cm]{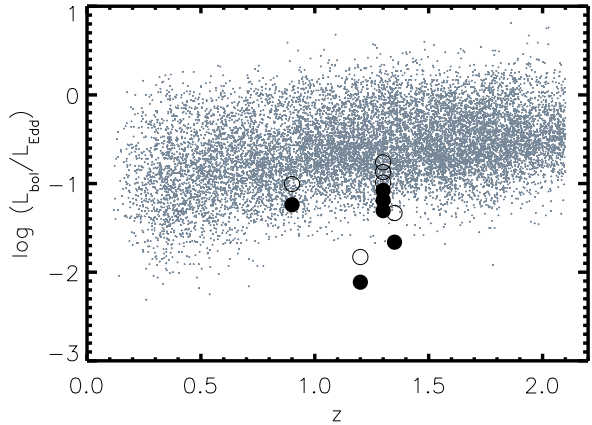} 
\caption{Bolometric luminosity as a fraction of the Eddington luminosity vs. redshift 
         for the whole sample of SDSS quasars of \citet{2004MNRAS.352.1390M}, plotted as 
         small crosses. The large circles indicate the six HELLAS2XMM AGN hosted in elliptical 
         galaxies (symbols as in Fig. \ref{ledd_fig}).}
 \label{mclure_dunlop}
\end{figure}

The estimated radiating efficiencies are significantly lower than the average 
$L_{bol}/L_{Edd}\approx$~0.4 inferred by \citet{2004MNRAS.351..169M}. 
However, since 
in the \citet{2004MNRAS.351..169M} model only the phases of significant black hole growth 
are considered, our results are not in contrast with the proposed model but suggest that our targets 
belong to the tail of the sources characterized by low accretion rates. 
Consistently, our data (black filled and empty circles representing the evolution 
and no-evolution hypothesis, respectively) lie in the lower envelope of the Eddington ratio 
distribution found by \citet{2004MNRAS.352.1390M} for their large SDSS quasar
sample. This is shown in Fig.~\ref{mclure_dunlop}, where our data are overlaid
on the SDSS data points.
This suggests that the SDSS quasar survey and the HELLAS survey
  probe different regimes of AGN activity: the SDSS samples the
  brightest sources in the sky ($R\lsimeq$20), most likely characterized by a high accretion
  rate, while our targets  (X-ray selected, optically faint, i.e., $R>24$, and obscured),
  are associated with a different evolutionary phase.
  We argue that the SMBH in our targets has already reached its final mass
  and the observed emission is witnessing a late stage of the accretion activity.



\section{Conclusions}
 
We have performed with \spitzer\ a pilot program to study a sample 
of eight Type~2 (i.e., luminous and obscured) quasars at high redshift, 
selected from the HELLAS2XMM survey. Three sources 
have a measured spectroscopic redshift (two secure and one tentative) from 
near-IR spectroscopy; the remaining objects have an estimated minimum redshift 
obtained from the $R-K$ colours. 
On the basis of their $K_{s}$-band morphological properties, the sample is divided into two classes: 
sources with radial profiles typical of elliptical galaxies and point-like objects. 
The most important results can be summarized as follows: 


\begin{itemize}

\item[$\bullet$]{All of the eight sources have been clearly detected in both IRAC and 
          MIPS 24~$\mu$m bands.}

\item[$\bullet$]{The \spitzer\ observations have allowed us to detect the nuclear 
     component (often hidden at short wavelengths by the host galaxy) as thermal 
     IR re-processed emission from the circumnuclear torus. 
     While for the two point-like sources the nuclear component dominates at all frequencies, 
     for the six sources with elliptical-like radial profile the contribution from the strong 
     stellar continuum is dominant up to the first IRAC bands, but the torus emission accounts 
     for the entire emission at 24~$\mu$m.}
 
\item[$\bullet$]{Taking advantage of the new \spitzer\ data, the nuclear SEDs of the sources 
     have been modeled and new photometric redshifts have been estimated, following two 
     approaches: for the elliptical sources, the nuclear emission has 
     been ``cleaned'' from the host galaxy contribution adopting a two-component model 
     (galaxy plus nuclear component), constrained using all the extensive observed data sets. 
     For the point-like sources, the SEDs appear inconsistent with an extinguished Type~1 quasar 
     template, being well reproduced by an empirical SED of red quasars \citep{2006ApJ...642..673P}. 
     We find an overall agreement between the SED templates 
     and the data points, and the derived photometric redshifts are consistent with the 
     spectroscopic ones for two sources.}

\item[$\bullet$]{Using the model components to extrapolate the nuclear SEDs in the far-IR regime, 
     we derived the bolometric luminosities (being in the range $\approx$~$10^{45}-10^{47}$~\lum) 
     by adding the IR luminosities to the full range of \xray\ luminosities. 
     In this computation, we have considered and discussed the 
     corrections to be applied to the observed IR luminosities to take into account
     the covering factor of the torus and the anisotropy of the IR 
     emission. The median 2--10~keV bolometric correction is $\approx$~25, 
     consistent with the value typically assumed in literature.} 

\item[$\bullet$]{For the elliptical sources, thanks to the independent estimates of the stellar
     light and nuclear bolometric luminosity, the physical parameters of the 
     central black holes have been estimated using the $M_{BH}-L_{K}$ relation and 
     exploring different hypotheses for the evolution of the $M_{BH}/M_{star}$ ratio with 
     redshift. 
     Under the hyphothesis that the $M_{BH}/M_{star}$ ratio is a factor of two higher at 
     $z\approx$~1.2 than locally, our luminous, obscured AGN have masses in the range 
     (0.2-2.5)${\times}10^{9}$~$M_{\odot}$, reside in massive 
     [(0.8-6.2)${\times}10^{11}$~$M_{\odot}$] high-redshift ellipticals 
     and are characterized by low Eddington ratios (${\lambda}{\approx}$~0.008--0.084). 
     Through our direct estimate of the IR luminosity, we confirm the conclusion of 
     \cite{2006A&A...445..457M} that these black holes may have already passed their 
     rapidly accretion phase}.

\end{itemize}

\begin{acknowledgements}
The authors acknowledge partial support by the Italian Space agency 
under the contract ASI--INAF I/023/05/0. The authors thank R. Gilli,
M. Polletta and L.Silva for useful discussions, 
and R.~J. McLure for kindly providing us with the data points of Fig.~\ref{mclure_dunlop}.
We thank the anonymous referee for the useful comments.
\end{acknowledgements}

\bibliographystyle{aa}
\bibliography{7092.bib}

\begin{thebibliography}{52}
\expandafter\ifx\csname natexlab\endcsname\relax\def\natexlab#1{#1}\fi

\bibitem[{{Antonucci}(1993)}]{1993ARA&A..31..473A}
{Antonucci}, R. 1993, \araa, 31, 473

\bibitem[{{Baldi} {et~al.}(2002){Baldi}, {Molendi}, {Comastri}, {Fiore},
  {Matt}, \& {Vignali}}]{2002ApJ...564..190B}
{Baldi}, A., {Molendi}, S., {Comastri}, A., {et~al.} 2002, \apj, 564, 190

\bibitem[{{Barger} {et~al.}(2005){Barger}, {Cowie}, {Mushotzky}, {Yang},
  {Wang}, {Steffen}, \& {Capak}}]{2005AJ....129..578B}
{Barger}, A.~J., {Cowie}, L.~L., {Mushotzky}, R.~F., {et~al.} 2005, \aj, 129,
  578

\bibitem[{{Brusa} {et~al.}(2005){Brusa}, {Comastri}, {Daddi}, {Pozzetti},
  {Zamorani}, {Vignali}, {Cimatti}, {Fiore}, {Mignoli}, {Ciliegi}, \&
  {R{\"o}ttgering}}]{2005A&A...432...69B}
{Brusa}, M., {Comastri}, A., {Daddi}, E., {et~al.} 2005, \aap, 432, 69

\bibitem[{{Bruzual} \& {Charlot}(2003)}]{2003MNRAS.344.1000B}
{Bruzual}, G. \& {Charlot}, S. 2003, \mnras, 344, 1000

\bibitem[{{Calzetti}(1997)}]{1997AIPC..408..403C}
{Calzetti}, D. 1997, in American Institute of Physics Conference Series, ed.
  W.~H. {Waller}, 403

\bibitem[{{Chabrier}(2003)}]{2003PASP..115..763C}
{Chabrier}, G. 2003, \pasp, 115, 763

\bibitem[{{Cocchia} {et~al.}(2007){Cocchia}, {Fiore}, {Vignali}, {Mignoli},
  {Brusa}, {Comastri}, {Feruglio}, {Baldi}, {Carangelo}, {Ciliegi}, {D'Elia},
  {La Franca}, {Maiolino}, {Matt}, {Molendi}, {Perola}, \&
  {Puccetti}}]{cocchia07}
{Cocchia}, F., {Fiore}, F., {Vignali}, C., {et~al.} 2007, \aap, in press,
  astro-ph/0612023

\bibitem[{{Comastri} \& {Fiore}(2004)}]{2004Ap&SS.294...63C}
{Comastri}, A. \& {Fiore}, F. 2004, \apss, 294, 63

\bibitem[{{Comastri} {et~al.}(1995){Comastri}, {Setti}, {Zamorani}, \&
  {Hasinger}}]{1995A&A...296....1C}
{Comastri}, A., {Setti}, G., {Zamorani}, G., \& {Hasinger}, G. 1995, \aap, 296,
  1

\bibitem[{{di Serego Alighieri} {et~al.}(2005){di Serego Alighieri}, {Vernet},
  {Cimatti}, {Lanzoni}, {Cassata}, {Ciotti}, {Daddi}, {Mignoli}, {Pignatelli},
  {Pozzetti}, {Renzini}, {Rettura}, \& {Zamorani}}]{2005A&A...442..125D}
{di Serego Alighieri}, S., {Vernet}, J., {Cimatti}, A., {et~al.} 2005, \aap,
  442, 125

\bibitem[{{Elvis} {et~al.}(1994){Elvis}, {Wilkes}, {McDowell}, {Green},
  {Bechtold}, {Willner}, {Oey}, {Polomski}, \& {Cutri}}]{1994ApJS...95....1E}
{Elvis}, M., {Wilkes}, B.~J., {McDowell}, J.~C., {et~al.} 1994, \apjs, 95, 1

\bibitem[{{Fabian}(2004)}]{2004cbhg.symp..446F}
{Fabian}, A.~C. 2004, in Coevolution of Black Holes and Galaxies, ed. L.~C.
  {Ho}, 446

\bibitem[{{Fadda} {et~al.}(2002){Fadda}, {Flores}, {Hasinger}, {Franceschini},
  {Altieri}, {Cesarsky}, {Elbaz}, \& {Ferrando}}]{2002A&A...383..838F}
{Fadda}, D., {Flores}, H., {Hasinger}, G., {et~al.} 2002, \aap, 383, 838

\bibitem[{{Fadda} {et~al.}(2006){Fadda}, {Marleau}, {Storrie-Lombardi},
  {Makovoz}, {Frayer}, {Appleton}, {Armus}, {Chapman}, {Choi}, {Fang},
  {Heinrichsen}, {Helou}, {Im}, {Lacy}, {Shupe}, {Soifer}, {Squires}, {Surace},
  {Teplitz}, {Wilson}, \& {Yan}}]{2006AJ....131.2859F}
{Fadda}, D., {Marleau}, F.~R., {Storrie-Lombardi}, L.~J., {et~al.} 2006, \aj,
  131, 2859

\bibitem[{{Fazio} {et~al.}(2004){Fazio}, {Hora}, {Allen}, {Ashby}, {Barmby},
  {Deutsch}, {Huang}, {Kleiner}, {Marengo}, {Megeath}, {Melnick}, {Pahre},
  {Patten}, {Polizotti}, {Smith}, {Taylor}, {Wang}, {Willner}, {Hoffmann},
  {Pipher}, {Forrest}, {McMurty}, {McCreight}, {McKelvey}, {McMurray}, {Koch},
  {Moseley}, {Arendt}, {Mentzell}, {Marx}, {Losch}, {Mayman}, {Eichhorn},
  {Krebs}, {Jhabvala}, {Gezari}, {Fixsen}, {Flores}, {Shakoorzadeh}, {Jungo},
  {Hakun}, {Workman}, {Karpati}, {Kichak}, {Whitley}, {Mann}, {Tollestrup},
  {Eisenhardt}, {Stern}, {Gorjian}, {Bhattacharya}, {Carey}, {Nelson},
  {Glaccum}, {Lacy}, {Lowrance}, {Laine}, {Reach}, {Stauffer}, {Surace},
  {Wilson}, {Wright}, {Hoffman}, {Domingo}, \& {Cohen}}]{2004ApJS..154...10F}
{Fazio}, G.~G., {Hora}, J.~L., {Allen}, L.~E., {et~al.} 2004, \apjs, 154, 10

\bibitem[{{Fiore} {et~al.}(2003){Fiore}, {Brusa}, {Cocchia}, {Baldi},
  {Carangelo}, {Ciliegi}, {Comastri}, {La Franca}, {Maiolino}, {Matt},
  {Molendi}, {Mignoli}, {Perola}, {Severgnini}, \&
  {Vignali}}]{2003A&A...409...79F}
{Fiore}, F., {Brusa}, M., {Cocchia}, F., {et~al.} 2003, \aap, 409, 79

\bibitem[{{Franceschini} {et~al.}(2005){Franceschini}, {Manners}, {Polletta},
  {Lonsdale}, {Gonzalez-Solares}, {Surace}, {Shupe}, {Fang}, {Xu}, {Farrah},
  {Berta}, {Rodighiero}, {Perez-Fournon}, {Hatziminaoglou}, {Smith}, {Siana},
  {Rowan-Robinson}, {Nandra}, {Babbedge}, {Vaccari}, {Oliver}, {Wilkes},
  {Owen}, {Padgett}, {Frayer}, {Jarrett}, {Masci}, {Stacey}, {Almaini},
  {McMahon}, {Johnson}, {Lawrence}, \& {Willott}}]{2005AJ....129.2074F}
{Franceschini}, A., {Manners}, J., {Polletta}, M.~d.~C., {et~al.} 2005, \aj,
  129, 2074

\bibitem[{{Gilli} {et~al.}(2007){Gilli}, {Comastri}, \&
  {Hasinger}}]{2007A&A...463...79G}
{Gilli}, R., {Comastri}, A., \& {Hasinger}, G. 2007, \aap, 463, 79

\bibitem[{{Granato} \& {Danese}(1994)}]{1994MNRAS.268..235G}
{Granato}, G.~L. \& {Danese}, L. 1994, \mnras, 268, 235

\bibitem[{{Granato} {et~al.}(1997){Granato}, {Danese}, \&
  {Franceschini}}]{1997ApJ...486..147G}
{Granato}, G.~L., {Danese}, L., \& {Franceschini}, A. 1997, \apj, 486, 147

\bibitem[{{Gregg} {et~al.}(2002){Gregg}, {Lacy}, {White}, {Glikman}, {Helfand},
  {Becker}, \& {Brotherton}}]{2002ApJ...564..133G}
{Gregg}, M.~D., {Lacy}, M., {White}, R.~L., {et~al.} 2002, \apj, 564, 133

\bibitem[{{Hickox} \& {Markevitch}(2006)}]{2006ApJ...645...95H}
{Hickox}, R.~C. \& {Markevitch}, M. 2006, \apj, 645, 95

\bibitem[{{Hopkins} {et~al.}(2006){Hopkins}, {Hernquist}, {Cox}, {Di Matteo},
  {Robertson}, \& {Springel}}]{2006ApJS..163....1H}
{Hopkins}, P.~F., {Hernquist}, L., {Cox}, T.~J., {et~al.} 2006, \apjs, 163, 1

\bibitem[{{Lonsdale} {et~al.}(2004){Lonsdale}, {Polletta}, {Surace}, {Shupe},
  {Fang}, {Xu}, {Smith}, {Siana}, {Rowan-Robinson}, {Babbedge}, {Oliver},
  {Pozzi}, {Davoodi}, {Owen}, {Padgett}, {Frayer}, {Jarrett}, {Masci},
  {O'Linger}, {Conrow}, {Farrah}, {Morrison}, {Gautier}, {Franceschini},
  {Berta}, {Perez-Fournon}, {Hatziminaoglou}, {Afonso-Luis}, {Dole}, {Stacey},
  {Serjeant}, {Pierre}, {Griffin}, \& {Puetter}}]{2004ApJS..154...54L}
{Lonsdale}, C., {Polletta}, M.~d.~C., {Surace}, J., {et~al.} 2004, \apjs, 154,
  54

\bibitem[{{Maiolino} {et~al.}(2006){Maiolino}, {Mignoli}, {Pozzetti},
  {Severgnini}, {Brusa}, {Vignali}, {Puccetti}, {Ciliegi}, {Cocchia},
  {Comastri}, {Fiore}, {La Franca}, {Matt}, {Molendi}, \&
  {Perola}}]{2006A&A...445..457M}
{Maiolino}, R., {Mignoli}, M., {Pozzetti}, L., {et~al.} 2006, \aap, 445, 457

\bibitem[{{Makovoz} \& {Marleau}(2005)}]{2005PASP..117.1113M}
{Makovoz}, D. \& {Marleau}, F.~R. 2005, \pasp, 117, 1113

\bibitem[{{Marconi} \& {Hunt}(2003)}]{2003ApJ...589L..21M}
{Marconi}, A. \& {Hunt}, L.~K. 2003, \apjl, 589, L21

\bibitem[{{Marconi} {et~al.}(2004){Marconi}, {Risaliti}, {Gilli}, {Hunt},
  {Maiolino}, \& {Salvati}}]{2004MNRAS.351..169M}
{Marconi}, A., {Risaliti}, G., {Gilli}, R., {et~al.} 2004, \mnras, 351, 169

\bibitem[{{McLure} \& {Dunlop}(2004)}]{2004MNRAS.352.1390M}
{McLure}, R.~J. \& {Dunlop}, J.~S. 2004, \mnras, 352, 1390

\bibitem[{{Mignoli} {et~al.}(2004){Mignoli}, {Pozzetti}, {Comastri}, {Brusa},
  {Ciliegi}, {Cocchia}, {Fiore}, {La Franca}, {Maiolino}, {Matt}, {Molendi},
  {Perola}, {Puccetti}, {Severgnini}, \& {Vignali}}]{2004A&A...418..827M}
{Mignoli}, M., {Pozzetti}, L., {Comastri}, A., {et~al.} 2004, \aap, 418, 827

\bibitem[{{Pei}(1992)}]{1992ApJ...395..130P}
{Pei}, Y.~C. 1992, \apj, 395, 130

\bibitem[{{Peng} {et~al.}(2006){Peng}, {Impey}, {Rix}, {Kochanek}, {Keeton},
  {Falco}, {Leh{\'a}r}, \& {McLeod}}]{2006ApJ...649..616P}
{Peng}, C.~Y., {Impey}, C.~D., {Rix}, H.-W., {et~al.} 2006, \apj, 649, 616

\bibitem[{{Perola} {et~al.}(2004){Perola}, {Puccetti}, {Fiore}, {Sacchi},
  {Brusa}, {Cocchia}, {Baldi}, {Carangelo}, {Ciliegi}, {Comastri}, {La Franca},
  {Maiolino}, {Matt}, {Mignoli}, {Molendi}, \& {Vignali}}]{2004A&A...421..491P}
{Perola}, G.~C., {Puccetti}, S., {Fiore}, F., {et~al.} 2004, \aap, 421, 491

\bibitem[{{Pier} \& {Krolik}(1993)}]{1993ApJ...418..673P}
{Pier}, E.~A. \& {Krolik}, J.~H. 1993, \apj, 418, 673

\bibitem[{{Polletta} {et~al.}(2006){Polletta}, {Wilkes}, {Siana}, {Lonsdale},
  {Kilgard}, {Smith}, {Kim}, {Owen}, {Efstathiou}, {Jarrett}, {Stacey},
  {Franceschini}, {Rowan-Robinson}, {Babbedge}, {Berta}, {Fang}, {Farrah},
  {Gonz{\'a}lez-Solares}, {Morrison}, {Surace}, \&
  {Shupe}}]{2006ApJ...642..673P}
{Polletta}, M.~d.~C., {Wilkes}, B.~J., {Siana}, B., {et~al.} 2006, \apj, 642,
  673

\bibitem[{{Richards} {et~al.}(2003){Richards}, {Hall}, {Vanden Berk},
  {Strauss}, {Schneider}, {Weinstein}, {Reichard}, {York}, {Knapp}, {Fan},
  {Ivezi{\'c}}, {Brinkmann}, {Budav{\'a}ri}, {Csabai}, \&
  {Nichol}}]{2003AJ....126.1131R}
{Richards}, G.~T., {Hall}, P.~B., {Vanden Berk}, D.~E., {et~al.} 2003, \aj,
  126, 1131

\bibitem[{{Richards} {et~al.}(2006){Richards}, {Lacy}, {Storrie-Lombardi},
  {Hall}, {Gallagher}, {Hines}, {Fan}, {Papovich}, {Vanden Berk}, {Trammell},
  {Schneider}, {Vestergaard}, {York}, {Jester}, {Anderson}, {Budav{\'a}ri}, \&
  {Szalay}}]{2006ApJS..166..470R}
{Richards}, G.~T., {Lacy}, M., {Storrie-Lombardi}, L.~J., {et~al.} 2006, \apjs,
  166, 470

\bibitem[{{Rieke} {et~al.}(2004){Rieke}, {Young}, {Engelbracht}, {Kelly},
  {Low}, {Haller}, {Beeman}, {Gordon}, {Stansberry}, {Misselt}, {Cadien},
  {Morrison}, {Rivlis}, {Latter}, {Noriega-Crespo}, {Padgett}, {Stapelfeldt},
  {Hines}, {Egami}, {Muzerolle}, {Alonso-Herrero}, {Blaylock}, {Dole}, {Hinz},
  {Le Floc'h}, {Papovich}, {P{\'e}rez-Gonz{\'a}lez}, {Smith}, {Su}, {Bennett},
  {Frayer}, {Henderson}, {Lu}, {Masci}, {Pesenson}, {Rebull}, {Rho}, {Keene},
  {Stolovy}, {Wachter}, {Wheaton}, {Werner}, \&
  {Richards}}]{2004ApJS..154...25R}
{Rieke}, G.~H., {Young}, E.~T., {Engelbracht}, C.~W., {et~al.} 2004, \apjs,
  154, 25

\bibitem[{{Rigby} {et~al.}(2004){Rigby}, {Rieke}, {Maiolino}, {Gilli},
  {Papovich}, {P{\'e}rez-Gonz{\'a}lez}, {Alonso-Herrero}, {Le Floc'h},
  {Engelbracht}, {Gordon}, {Hines}, {Hinz}, {Morrison}, {Muzerolle}, {Rieke},
  \& {Su}}]{2004ApJS..154..160R}
{Rigby}, J.~R., {Rieke}, G.~H., {Maiolino}, R., {et~al.} 2004, \apjs, 154, 160

\bibitem[{{Salpeter}(1955)}]{1955ApJ...121..161S}
{Salpeter}, E.~E. 1955, \apj, 121, 161

\bibitem[{{Schmidt} \& {Green}(1983)}]{1983ApJ...269..352S}
{Schmidt}, M. \& {Green}, R.~F. 1983, \apj, 269, 352

\bibitem[{{Shields} {et~al.}(2006){Shields}, {Salviander}, \&
  {Bonning}}]{2006NewAR..50..809S}
{Shields}, G.~A., {Salviander}, S., \& {Bonning}, E.~W. 2006, New Astronomy
  Review, 50, 809

\bibitem[{{Silva} {et~al.}(2004){Silva}, {Maiolino}, \&
  {Granato}}]{2004MNRAS.355..973S}
{Silva}, L., {Maiolino}, R., \& {Granato}, G.~L. 2004, \mnras, 355, 973

\bibitem[{{Soltan}(1982)}]{1982MNRAS.200..115S}
{Soltan}, A. 1982, \mnras, 200, 115

\bibitem[{{Spergel} {et~al.}(2003){Spergel}, {Verde}, {Peiris}, {Komatsu},
  {Nolta}, {Bennett}, {Halpern}, {Hinshaw}, {Jarosik}, {Kogut}, {Limon},
  {Meyer}, {Page}, {Tucker}, {Weiland}, {Wollack}, \&
  {Wright}}]{2003ApJS..148..175S}
{Spergel}, D.~N., {Verde}, L., {Peiris}, H.~V., {et~al.} 2003, \apjs, 148, 175

\bibitem[{{Surace} {et~al.}(2005){Surace}, {Shupe}, {Fang}, \& {et
  al.}}]{surace05}
{Surace}, J.~A., {Shupe}, D.~L., {Fang}, F., \& {et al.} 2005, technical
  report, The SWIRE Data Release 2. Available at
  \verb+http://swire.ipac.caltech.edu/swire/astronomers/+
  \verb+publications/SWIRE2_doc_083105.pdf+

\bibitem[{{Vignali} {et~al.}(2003){Vignali}, {Brandt}, \&
  {Schneider}}]{2003AJ....125..433V}
{Vignali}, C., {Brandt}, W.~N., \& {Schneider}, D.~P. 2003, \aj, 125, 433

\bibitem[{{Vignali} {et~al.}(2005){Vignali}, {Brandt}, {Schneider}, \&
  {Kaspi}}]{2005AJ....129.2519V}
{Vignali}, C., {Brandt}, W.~N., {Schneider}, D.~P., \& {Kaspi}, S. 2005, \aj,
  129, 2519

\bibitem[{{Werner} {et~al.}(2004){Werner}, {Roellig}, {Low}, {Rieke}, {Rieke},
  {Hoffmann}, {Young}, {Houck}, {Brandl}, {Fazio}, {Hora}, {Gehrz}, {Helou},
  {Soifer}, {Stauffer}, {Keene}, {Eisenhardt}, {Gallagher}, {Gautier}, {Irace},
  {Lawrence}, {Simmons}, {Van Cleve}, {Jura}, {Wright}, \&
  {Cruikshank}}]{2004ApJS..154....1W}
{Werner}, M.~W., {Roellig}, T.~L., {Low}, F.~J., {et~al.} 2004, \apjs, 154, 1

\bibitem[{{Woo} {et~al.}(2006){Woo}, {Treu}, {Malkan}, \&
  {Blandford}}]{2006ApJ...645..900W}
{Woo}, J.-H., {Treu}, T., {Malkan}, M.~A., \& {Blandford}, R.~D. 2006, \apj,
  645, 900

\bibitem[{{Zamorani} {et~al.}(1981){Zamorani}, {Henry}, {Maccacaro},
  {Tananbaum}, {Soltan}, {Avni}, {Liebert}, {Stocke}, {Strittmatter},
  {Weymann}, {Smith}, \& {Condon}}]{1981ApJ...245..357Z}
{Zamorani}, G., {Henry}, J.~P., {Maccacaro}, T., {et~al.} 1981, \apj, 245, 357

\end{thebibliography}

\end{document}